\documentclass[epjST]{svjour}
\usepackage{graphicx}

\usepackage{mathptmx, courier, pifont}
\usepackage[scaled=0.92]{helvet}
\usepackage[T1]{fontenc}
\usepackage{textcomp} 

\usepackage{amsmath} 

\usepackage{bm}

\newcommand{\fig}[1]{Fig.~\ref{fig:#1}}

\newcommand{\su}[1]{\mbox{SU($#1$)}}
\newcommand{\so}[1]{\mbox{SO($#1$)}}

\newcommand{\tsub}[1]{_{\mbox{\scriptsize#1}}}

\newcommand{\sideparg}[1]{\medskip\noindent{\bf\em #1}:}

\newcommand{\singlefig}[6]{%
\begin{figure}\vspace{#3}%
\includegraphics*[scale=#5]{#2}%
\caption{\label{fig:#1} #6}%
\vspace{#4}%
\end{figure}}

\begin{document}

\title{The Superconducting Critical Temperature}

\author{
Mike Guidry\inst{1}\fnmsep\thanks{\email{guidry@utk.edu}} 
\and 
Yang Sun\inst{2}\fnmsep\thanks{\email{sunyang@sjtu.edu.cn}}
\and 
Lian-Ao Wu \inst{3}\fnmsep\thanks{\email{lianaowu@gmail.com}} 
}
\institute{
Department of Physics and Astronomy, University of Tennessee, 
Knoxville TN 37996, USA 
\and
School of Physics and Astronomy, Shanghai Jiao Tong University, Shanghai
200240, People's Republic of China 
\and
IKERBASQUE, Basque Foundation for Science, 48011 Bilbao, Spain,
and Department of Theoretical Physics and History of Science,
Basque Country University (EHU/UPV), Post Office Box 644, 48080 Bilbao, Spain
}
\abstract{
Two principles govern the critical temperature for superconducting transitions: 
(1)~intrinsic strength of the pair coupling and (2)~effect of the many-body 
environment on the efficiency of that coupling. Most discussions take into 
account only the first but we argue that the properties of unconventional 
superconductors are governed more often by the second, through dynamical 
symmetry relating  normal and superconducting states. Differentiating these 
effects is essential to charting a path to the highest-temperature 
superconductors.
}
\maketitle

\section{Introduction}
\label{intro}

Conventional superconductivity (SC) is described well by BCS theory using 
spherical ($s$-wave) pairing formfactors corresponding to phonon pair binding.  
Superconductors with formfactors that are not $s$-wave are termed {\em 
unconventional}; the most famous are  cuprate high-temperature superconductors, 
which have $d$-wave pairing  and  high transition temperatures $T\tsub c$ 
relative to conventional SC. Many other unconventional superconductors are 
known, often exhibiting  $T_{\rm c}$  larger than for conventional SC.  
Superconductors involve Cooper-pair condensates  and stronger pairing favors 
survival of the condensate at higher $T$.  Thus, enhancing phonon coupling by 
tuning atomic mass and lattice spacing can increase $T\tsub c$ for conventional 
SC. For unconventional SC the situation is more nuanced.  The value of $T\tsub 
c$ depends on intrinsic pairing strength (through  electron correlations rather 
than phonons), but there is another factor, often more important and often 
overlooked in standard discussions.  This is seen most clearly by viewing the SC 
transition from the perspective of {\em fermion dynamical symmetry.}

\section{Fermion Dynamical Symmetry and Superconductivity}
\label{sec:fdsmSC}

The fermion dynamical symmetry method uses principles of dynamical symmetry to 
truncate a Hilbert space to a collective subspace, specified in terms of a Lie 
algebra having a relatively small number of fermionic generators, as illustrated 
in \fig{dynamicalSymmMicroscopic}(a).%
\singlefig
{dynamicalSymmMicroscopic}
{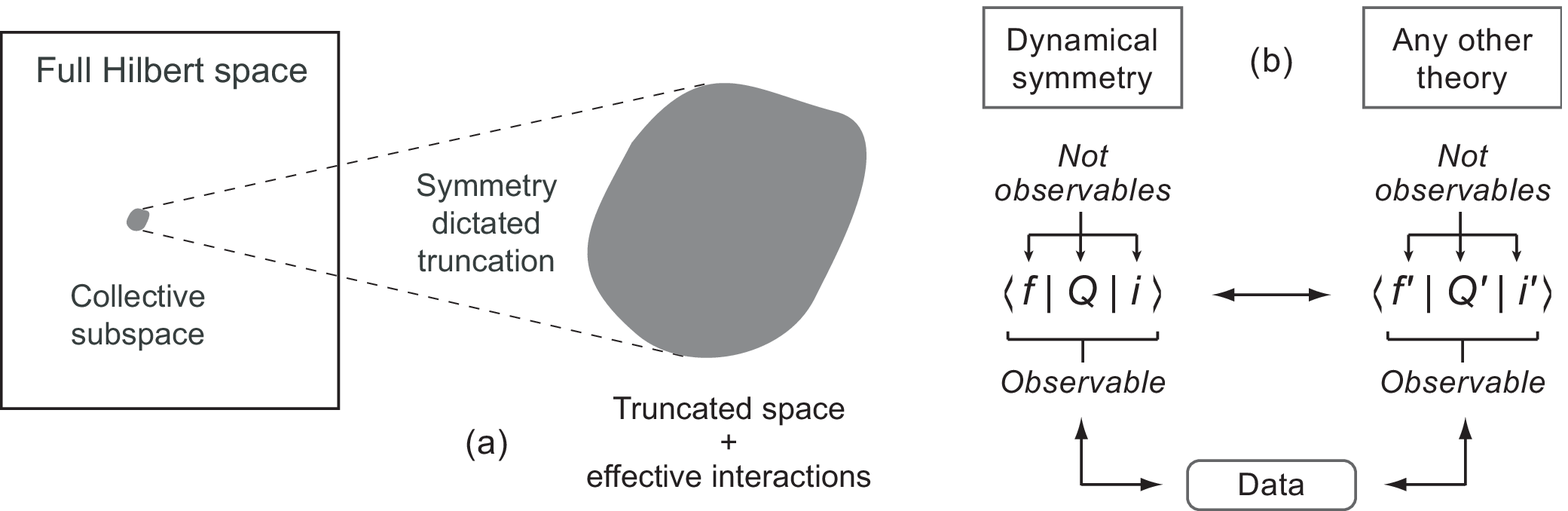}
{0pt}
{0pt}
{0.615}
{(a)~Emergent-symmetry truncation of Hilbert space to a collective subspace 
using principles of dynamical symmetry. (b)~Comparison of matrix elements among 
different theories and data. Wavefunctions and operators are {\em not 
observables}.  Only matrix elements are directly related to observables.}
Within the subspace the most general Hamiltonian is a polynomial in the Casimir 
invariants for all the subgroup chains of the highest symmetry consistent with 
conservation laws, with coefficients of the terms  determined by effective 
interactions representing the average effects of the space excluded by the 
truncation.  Matrix elements  can  be determined exactly in specific limits and 
approximately using  coherent state methods otherwise. As illustrated in 
\fig{dynamicalSymmMicroscopic}(b), the theory is {\em microscopic} because the 
only valid comparison in quantum mechanics is for matrix elements.

A description of cuprate superconductors by this method  is outlined in Figs.\ 
\ref{fig:su4-so8Competition} and \ref{fig:su4-so8_relationship}.%
\singlefig
     {su4-so8Competition}
     {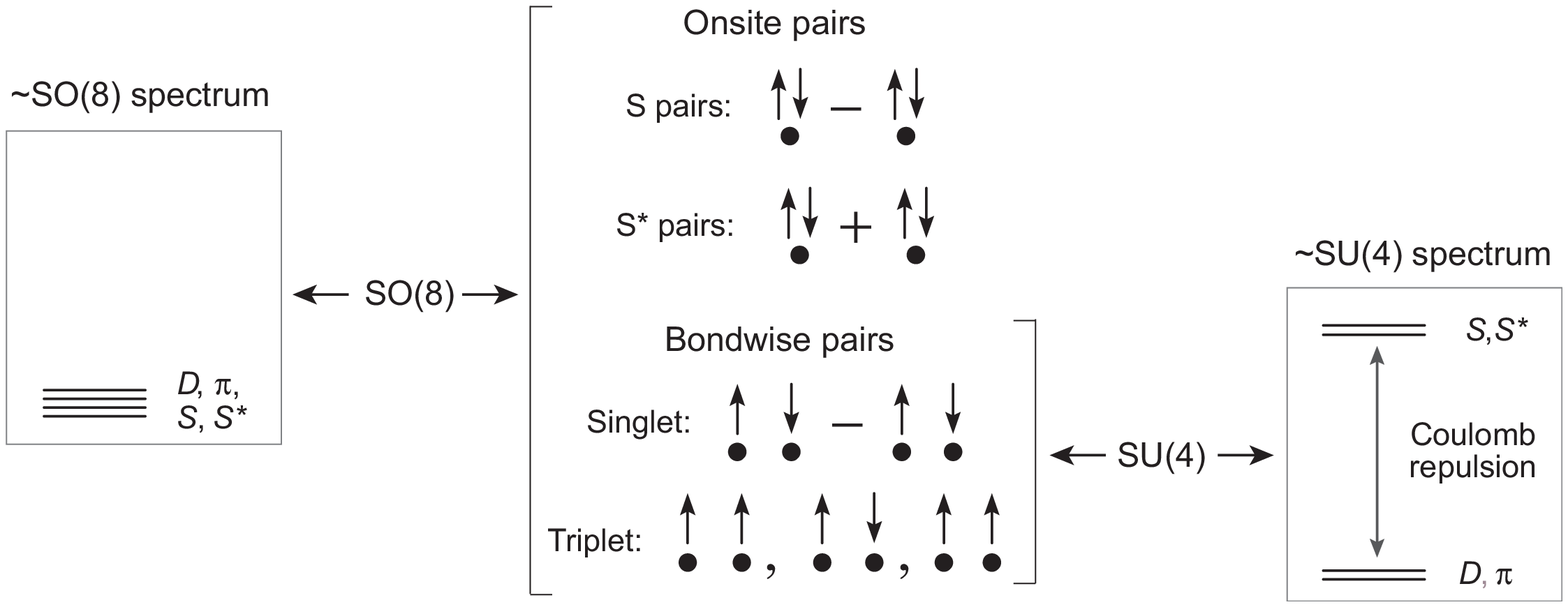}
     {0pt}
     {0pt}
     {0.57}
     {Schematic difference between bondwise $(D, \pi)$ and onsite $(S, S^*)$ 
pair energies.    If onsite repulsion is weak the pairing states are nearly 
degenerate, yielding an SO(8) symmetry.  If it is strong onsite pairs are pushed 
up in energy, reducing the  symmetry to an effective SU(4)  low-energy 
symmetry.}
 \singlefig
{su4-so8_relationship} 
{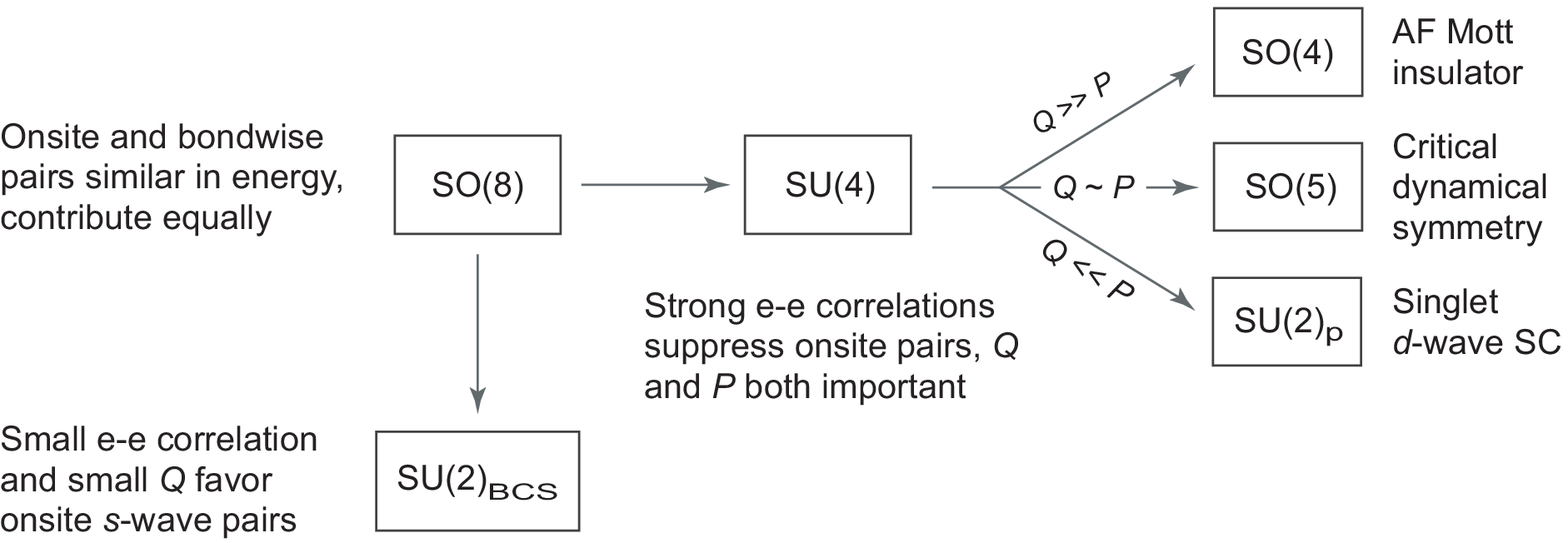}
{0pt}
{0pt}
{0.68}
{The relationship between SO(4), SO(8), and BCS SU(2) symmetry for 
conventional and unconventional superconductors.}
Restricting to the 28 generators describing onsite and nearest-neighbor pairing, 
antiferromagnetic, spin, and charge degrees of freedom, the minimal symmetry is 
\so8. If Coulomb repulsion is weak and antiferromagnetic (AF) correlations are 
negligible, onsite pairs are favored over bondwise pairs (for simplicity we 
restrict to nearest-neighbor bondwise pairs here). This favors subgroup chains 
containing \su2 pseudospin generators that give conventional SC described by 
$\su2_{\rm BCS}$ symmetry. If Coulomb repulsion is strong and 
antiferromagnetism significant, onsite pairing is disfavored relative to 
bondwise pairing and antiferromagnetic operators become important in addition to 
pairing operators.  This reduces \so8 to a 15-generator subgroup \su4, with 
generators representing AF, spin-singlet and spin-triplet bondwise pairs, spin, 
and charge operators; explicit forms for the operators and their 
commutation relations are given in Refs.\ \cite{guid2001,su4review}. Three 
dynamical symmetry chains  have exact solutions and correspond (through their 
matrix elements) to physical states thought to be relevant for  cuprate doped 
and undoped states:

\begin{enumerate}
 \item 
 SU(4) $\supset$ SO(4),  which  represents an antiferromagnetic (AF) Mott 
insulating state that  is the low-temperature ground state for zero doping.

\item
SU(4) $\supset$ SU(2),  which  represents a $d$-wave singlet superconducting 
(SC) state that can become the low-temperature ground state for non-zero doping.

\item
SU(4) $\supset$ SO(5),   which represents a critical dynamical symmetry 
interpolating between the SU(2) superconducting and SO(4) antiferromagnetic 
solutions.
\end{enumerate}

\noindent
We will now document concisely in Section \ref{sec:su4andData} that this 
microscopic approach  gives a remarkably good description of a broad range of 
cuprate phenomena with  minimal assumptions, and then use the validated \su4  
theory to discuss SC transition temperatures in Section 
\ref{sec:highTcunconventional}.

\section{SU(4) Dynamical Symmetry and Cuprate Phenomenology}
\label{sec:su4andData}

The properties of high-temperature superconductors (HTSC) raise fundamental 
questions that have proven difficult to answer in a comprehensive way: (1)~What 
physics controls the  phase diagram? (2)~What role do quantum phase transitions 
(QFT) and quantum critical behavior play, and what is their microscopic origin? 
(3)~How does the Cooper instability arise in a doped Mott insulator? (4)~What is 
the origin of the pseudogap (PG), and do ``competing order'' or ``preformed 
pairs'' govern its properties; how is the PG related to the AF and SC phases? 
(5)~Why do underdoped cuprates  exhibit complexity and disorder despite a highly 
universal overall phase diagram? (6)~How are HTSC (and other unconventional 
superconductors) related to conventional SC? (7)~Why is $T\tsub c$ higher than 
expected for  unconventional superconductors? (8)~What  principles can guide 
searches for new high-$T\tsub c$ superconductors? (9)~How is HTSC related to the 
various forms of SC and superfluidity that occur in other fields of physics? In 
our opinion, no standard  approach can provide plausible answers to this entire 
list without ad hoc assumptions. Let us now apply the the \su4 model of HTSC  to 
 answering the questions posed above. The model is documented in a series of 
publications 
\cite{guid2001,lawu03,guid04,sun05,sun06,sun07,guid09,guid09b,guid10,guid11} and 
a comprehensive review \cite{su4review}. Here we collect in one place a unified 
set of physical implications for \su4 symmetry, unobscured by technical detail. 
Hence, we shall write few equations, preferring to emphasize physical 
interpretation and referencing the literature where ample equations, 
derivations, and technical justification may be found.

\sideparg{Origin of the Phase Diagram}
Universality of the cuprate phase diagram suggests  a unifying principle 
independent of microscopic details.  The \su4 model implies that symmetry alone 
dictates many basic properties,  and that these properties lead to a highly 
universal phase diagram, illustrated in \fig{gapsResolve_k_BW},%
\singlefig
     {gapsResolve_k_BW}
     {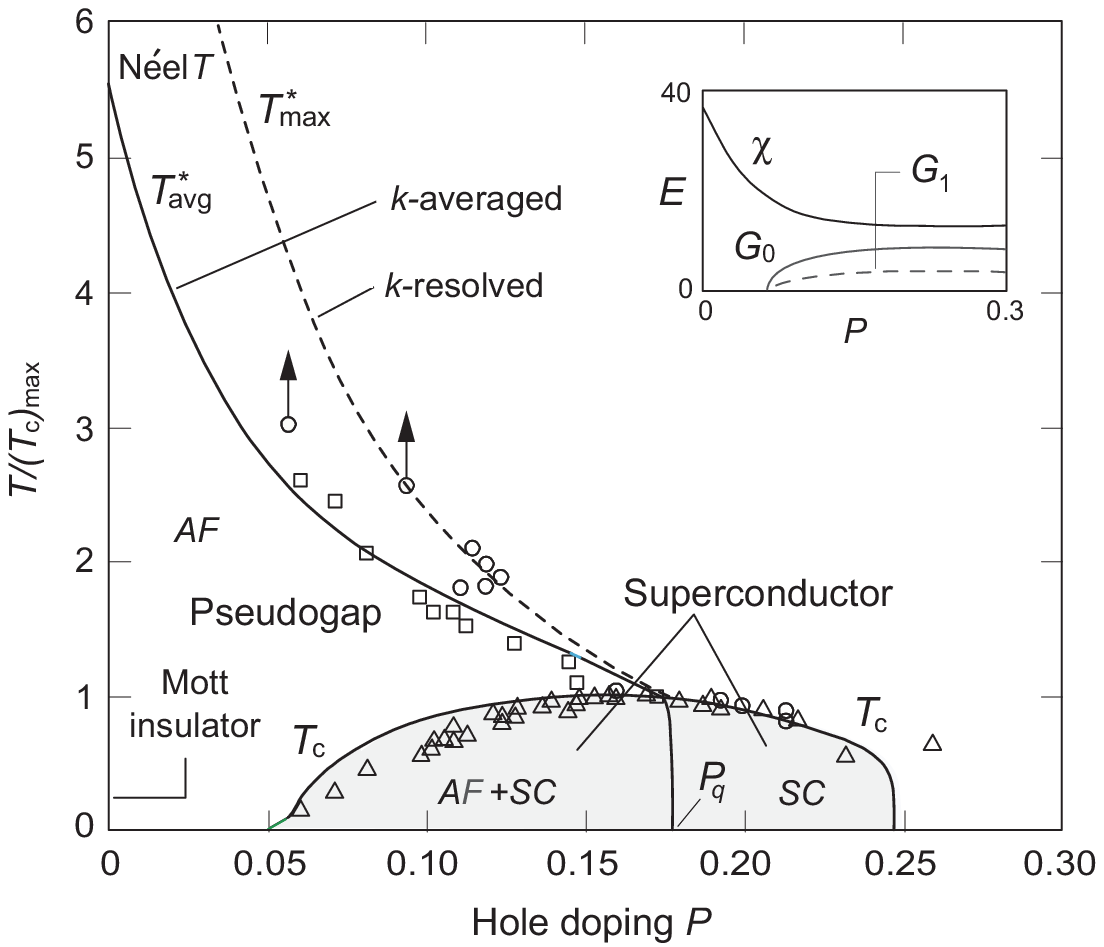}
     {0pt}
     {0pt}
     {0.76}
     {SU(4) cuprate temperature $T$ and doping $P$ phase diagram compared with 
data taken from Refs.\ \cite{dai99,camp99}. Strengths of the AF and singlet 
pairing correlations were determined in Ref.\ \cite{sun06} by global fits to 
cuprate data (inset plot). Pseudogap temperatures are indicated by $T^*$.  The 
two PG curves correspond to whether momentum is resolved or not in the 
experiment. The inset shows the variation of the AF and pairing coupling with 
doping $P$.}
that is described quantitatively by the \su4 model.  Only two significant 
parameters enter: the effective strength of singlet pairing $G_0$ and the 
effective strength of AF correlations $\chi$ (triplet pairing strength $G_1$ has 
minimal influence).  The best fit is for the smooth dependence of 
$G_0$ and $\chi$ on the doping $P$ shown in the inset of \fig{gapsResolve_k_BW}, 
but the basic features survive if these parameters are held 
constant with doping (see Ref.\ \cite{su4review}).  Thus {\em the 
cuprate phase diagram is a consequence of  SU(4) symmetry} correlating 
emergent $d$-wave singlet pairing and antiferromagnetism; it {\em depends only 
parametrically microscopic details} such as pairing formfactors.

\sideparg{AF Mott Insulator States at Half Filling}
By counting,  \su4 symmetry requires {\em no double 
occupancy of lattice sites}  by correlated fermions \cite{guid04}.  Hence, 
charge transport is suppressed at half band-filling and the undoped 
ground state is a {\em Mott insulator.}   Moreover,  this state has 
$\su4\supset \so4$ dynamical symmetry and the matrix elements of an AF 
N\'eel state \cite{guid2001,lawu03,su4review}. Thus, the undoped \su4  ground 
state is an {\em AF Mott insulator}, just as observed for cuprates.

\sideparg{Cooper Instability of the Doped Mott Insulator}
The same \su4 symmetry requiring the undoped ground state to be an AF Mott 
insulator implies that this state is {\em fundamentally unstable against 
condensing Cooper pairs} when doped \cite{guid10,su4review}.  This results in a 
quantum phase transition (QPT) to be discussed more extensively below, and  
implies a rapid transition to a superconducting state upon doping, as observed 
for data in  \fig{gapsResolve_k_BW}.

\sideparg{Upper Doping Limit for Superconductivity}
Direct counting implies that occupancy of more than $\frac14$ of 
lattice sites by  holes will break  \su4 and destroy 
SC \cite{guid2001,guid04}.  This in accord with the data displayed in 
\fig{gapsResolve_k_BW}, where $T\tsub c > 0$ only for doping less than 
$\sim25\%$ holes.

\sideparg{Optimal Doping for  Superconductivity}
 For doping larger than that near the peak of the superconducting dome (optimal 
doping) in \fig{gapsResolve_k_BW}, SC properties are observed to become better 
defined. As discussed further below, this  is a natural consequence of \su4 
symmetry, which implies  a quantum phase transition  near optimal doping 
exhibiting critical behavior \cite{su4review}. At subcritical doping  the 
superconducting wavefunction is perturbed by residual AF correlations. At the 
QFT the AF correlations vanish identically, leaving  pure $d$-wave, BCS-like, 
singlet SC  above  critical doping. This is consistent with various cuprate 
experiments.

\sideparg{Existence of a Pseudogap}
A  pseudogap is  a partial gap at the Fermi surface above  $T\tsub c$.  From 
\fig{gapsResolve_k_BW}, at lower doping it is the ``normal state'' from which  
SC can be produced by lowering the temperature through the doping-dependent 
critical temperature $T\tsub c$.  As explained further below, a PG is expected 
from AF--SC competition in the  \su4 Hilbert subspace \cite{su4review}.

\sideparg{Quantum Critical Behavior}
A  highest symmetry with multiple dynamical symmetry subchains leads naturally 
to quantum phase transitions as tuning parameters such as doping, magnetic 
field, or pressure shift the balance between competing dynamical symmetries.  
The \su4 theory is microscopic so one can determine whether these transitions 
are associated with critical behavior and examine the corresponding physical 
consequences.  Thus, \su4 and its dynamical symmetry subchains are a laboratory 
for quantum critical behavior in HTSC. As we now discuss, the \su4 model implies 
two fundamental instabilities leading to quantum phase transitions that are 
central to understanding HTSC, and a {\em critical dynamical symmetry} that 
generalizes  a quantum critical point to an entire {\em quantum critical phase}, 
which proves useful in understanding the underdoped region in general and the PG 
region in particular.

\sideparg{The SU(4) Cooper Instability}
The \su4 solution at  $T=0$ for the pairing order parameter $\Delta$ given  in 
Eq.~(24b) of Ref.~\cite{sun05} implies that \cite{guid10}
\begin{equation}
    \left. \frac{\partial\Delta}{\partial x} \right|_{x=0} =
        \left. \frac14 \frac{x_q^{-1} -2x}{(x(x_q^{-1} -x))^{1/2}}
\right|_{x=0}
        = \infty ,
\label{analyticalDelderiv}
\end{equation}
where $x$ is  doping  and $x_q$ is a critical doping value predicted by the 
theory. (In \fig{gapsResolve_k_BW} the critical doping point is labeled $P_q$ 
with $x_q \sim 4P_q$.)  This implies a fundamental pairing instability at $x=0$: 
 the \su4 AF Mott insulator ground state at half filling is intrinsically 
unstable against a QFT that condenses singlet hole pairs for infinitesimal hole 
doping in the presence of non-zero attractive pairing \cite{guid10}.  Figure 
\ref{fig:instabilities}(a) illustrates.%
\singlefig 
{instabilities} 
{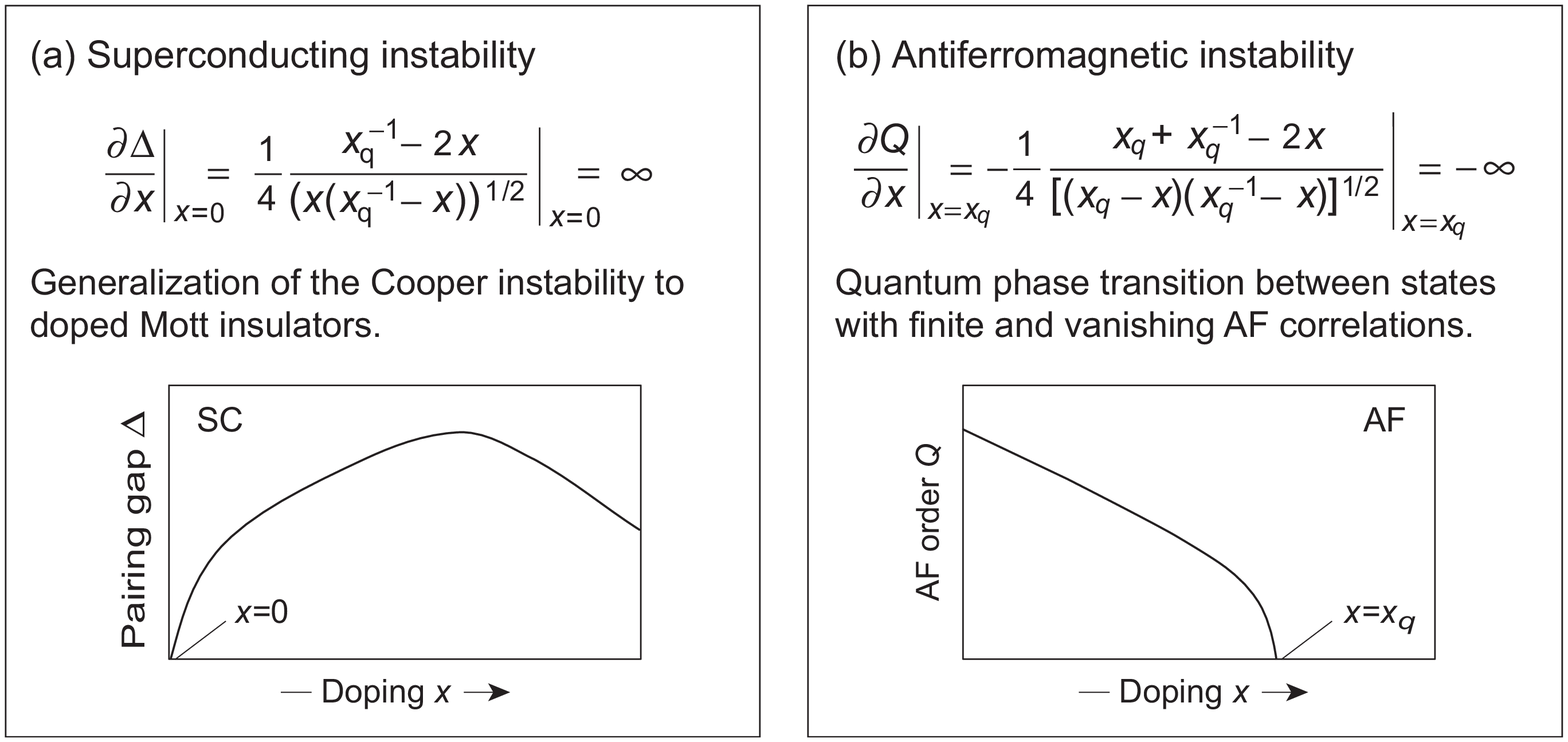}
{0pt} 
{0pt} 
{0.45}
{Two fundamental \su4 instabilities that govern the behavior of high temperature 
superconductors. The plots illustrate (a)~the generalized Cooper instability and 
(b)~The  AF instability in terms of the values of the order 
parameters calculated within coherent state approximation}
Hence, the rapid onset of SC with hole-doping in the cuprates results from a 
Cooper-like instability for $d$-wave pairs in an AF Mott insulator. The \su4
solution reduces to ordinary $d$-wave BCS theory if the AF interaction 
vanishes and to an AF Mott insulator if the pairing interaction vanishes 
~\cite{sun05}.  Thus it generalizes the Cooper instability  to 
doped Mott insulators and may be viewed as the traditional Cooper instability 
for a Fermi sea polarized by strong onsite Coulomb repulsion and AF 
correlations, or equivalently as a Fermi sea exhibiting \su4 symmetry
\cite{su4review}.

\sideparg{The SU(4) AF Instability}
\su4 symmetry implies a second fundamental instability.  From the \su4 solution 
for the AF order parameter $Q$ given by Eqs.~(24b, 14) of Ref.~\cite{sun05},
\begin{equation}
    \left. \frac{\partial Q}{\partial x} \right|_{x=x_q} =
        \left. -\frac14 \frac{x_q + x_q^{-1}-2x}
        {[(x_q-x)(x_q^{-1} -x)]^{1/2}} \right|_{x=x_q}
        = - \infty ,
\label{analyticalDelderiv2}
\end{equation}
and a small change in doping  causes a divergence in AF correlations  near the 
critical doping  $ x = x_q = 4P_q $ in \fig{gapsResolve_k_BW} 
\cite{guid11,su4review}.   Figure \ref{fig:instabilities}(b) illustrates.  This 
instability is associated with a QFT between a SC state still influenced by AF 
correlations and a pure SC state.

\sideparg{Dynamical Symmetries and Critical Behavior}
Quantum phase transitions and quantum critical points are  a natural consequence 
of fermion dynamical symmetries, implying that quantum critical behavior is a 
{\em corollary} of unconventional superconductivity, {\em not a cause.} 
Furthermore, some dynamical symmetry solutions generalize quantum critical 
points to entire {\em quantum critical phases} exhibiting critical behavior 
\cite{lawu03,guid11,su4review}.  The $\su4\supset\so5$ dynamical symmetry is an 
example.  This is seen most easily in generalized coherent state approximation 
\cite{lawu03}, which represents symmetry-constrained Hartree--Fock--Bogoliubov 
solutions that permit \su4 results to be expressed in terms of gap equations and 
quasiparticles, and that lead naturally to total energy surfaces connecting \su4 
solutions microscopically to Ginzburg--Landau theory.  \su4 coherent state 
energy surfaces are displayed in \fig{eSurfacesAndCompare}(a-c). The flat 
critical nature of the \so5 surface is evident for low doping  in 
\fig{eSurfacesAndCompare}(b).
\singlefig
{eSurfacesAndCompare}
{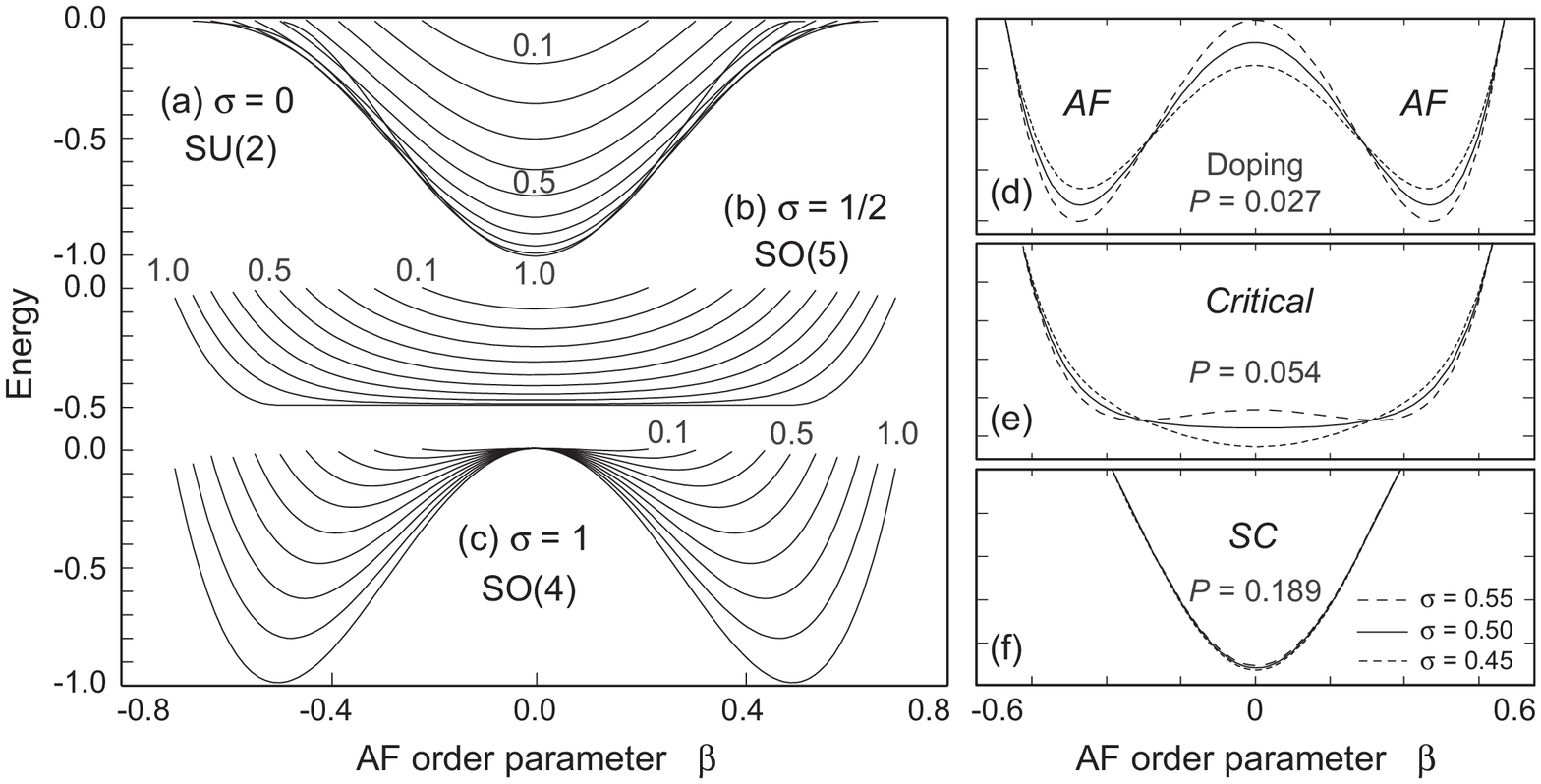}
{0pt}
{0pt}
{0.564}
{(a-c)~Coherent-state energy surfaces for symmetry limits of the \su4
Hamiltonian \cite{lawu03}. The horizontal axis measures  AF order. Curves are 
labeled by lattice occupation fractions with the value 1 corresponding to half 
filling. The parameter $\sigma$ is the ratio of AF coupling to the sum of AF and 
pairing coupling strengths. (d-f)~Effect of altering the ratio $\sigma$  for 
three values of doping in the cuprates. In (d) and (f) the system is in the 
stable minima associated with AF and SC, respectively, and changing $\sigma$ by 
10\% hardly alters the location of the energy minima, but in (b) the energy 
surface is critical and the perturbation can flip the  nature of the ground 
state between SC and AF minima.
}

\sideparg{Complexity in the Underdoped Region} 
As suggested in \fig{eSurfacesAndCompare}(d-f), the underdoped $\sim$\so5 energy 
surface exhibits {\em complexity} because many potential ground states with very 
different order parameters have almost the same energy.  Complexity implies 
susceptibility to  fluctuations in AF and SC order induced by small 
perturbations and the phase defined by the \so5 dynamical symmetry is a {\em 
critical dynamical symmetry.}  Critical dynamical symmetries are a fundamental 
organizing principle for complexity in strongly-correlated nuclear structure and 
condensed matter systems \cite{guid11,su4review}. In Ref.\ \cite{guid11} we 
have proposed that stripe and checkerboard patterns, amplification of proximity 
effects, and related phenomena common in underdoped compounds may be a 
consequence of complexity enabled by the critical nature of the energy surface 
there.  This complexity can occur with or without associated spatial modulation 
of charge. Charge is not an \su4 generator so it is not fundamental for HTSC, 
but it can play a secondary role by perturbing critical energy surfaces in 
underdoped compounds.

\sideparg{Competing Order or Preformed Pairs}
In the {\em competing-order model} the PG is an energy scale for order competing 
with SC. In the rival {\em preformed pairs model}  pairs form at higher energy 
with phase fluctuations that suppress long-range order until at a lower energy 
the pairs condense into a SC with long-range order. In \su4 the PG  scale  is an 
AF correlation  competing with SC but the AF operators are generators of \su4 
and the collective  subspace is a superposition of pairs.   Thus, the PG 
results from a superposition of \su4 pairs that can condense into a strong 
superconductor only after AF fluctuations are suppressed by doping. Hence the 
\su4 pseudogap state results from competing AF and SC order in a basis of 
fermion pairs and {\em SU(4)  unifies the competing order and preformed pair 
pictures.}

\sideparg{Fermi Arcs and Anisotropy of Pseudogaps}
The \su4 cuprate model implies strong angular dependence  in $k$-space, which 
leads to temperature and doping restrictions on regions of the Brillouin zone 
where ungapped Fermi surface can exist \cite{guid09b,su4review}.  If this region 
is interpreted in terms of Fermi arcs, the \su4 model gives a natural 
description of arc lengths as a function of temperature in quantitative accord 
with ARPES data \cite{guid09b,su4review}. If the Fermi surface is interpreted in 
terms of small pockets instead, \su4 symmetry restricts their possible location 
and size.

\sideparg{New Superconductors}
 In 2004 we argued that the essence of cuprate SC is non-abelian dynamical 
symmetry,  and that compounds with similar symmetries but different microscopic 
structure should exist \cite{guid04}. Discovery of Fe-based SC in 2008 validated 
this prediction \cite{guid09,su4review}.

\sideparg{SU(4) and Conventional BCS Superconductivity}
The relationship of  \su4  to  BCS SC was given in \fig{su4-so8_relationship}. 
Conventional SC is the limit of  $\so8 \supset \su4$ SC  when Coulomb repulsion 
is small and AF correlation is negligible. Thus $\so8 \supset \su4$ dynamical 
symmetry provides a unified view of conventional and unconventional 
superconductivity \cite{su4review}.

\sideparg{Dynamical Symmetry and Universality of Emergent States}
Dynamical symmetries for a variety of emergent states suggest an even 
broader unification transcending fields and subfields.  Figure 
\ref{fig:phaseComposite}%
\singlefig
     {phaseComposite}
     {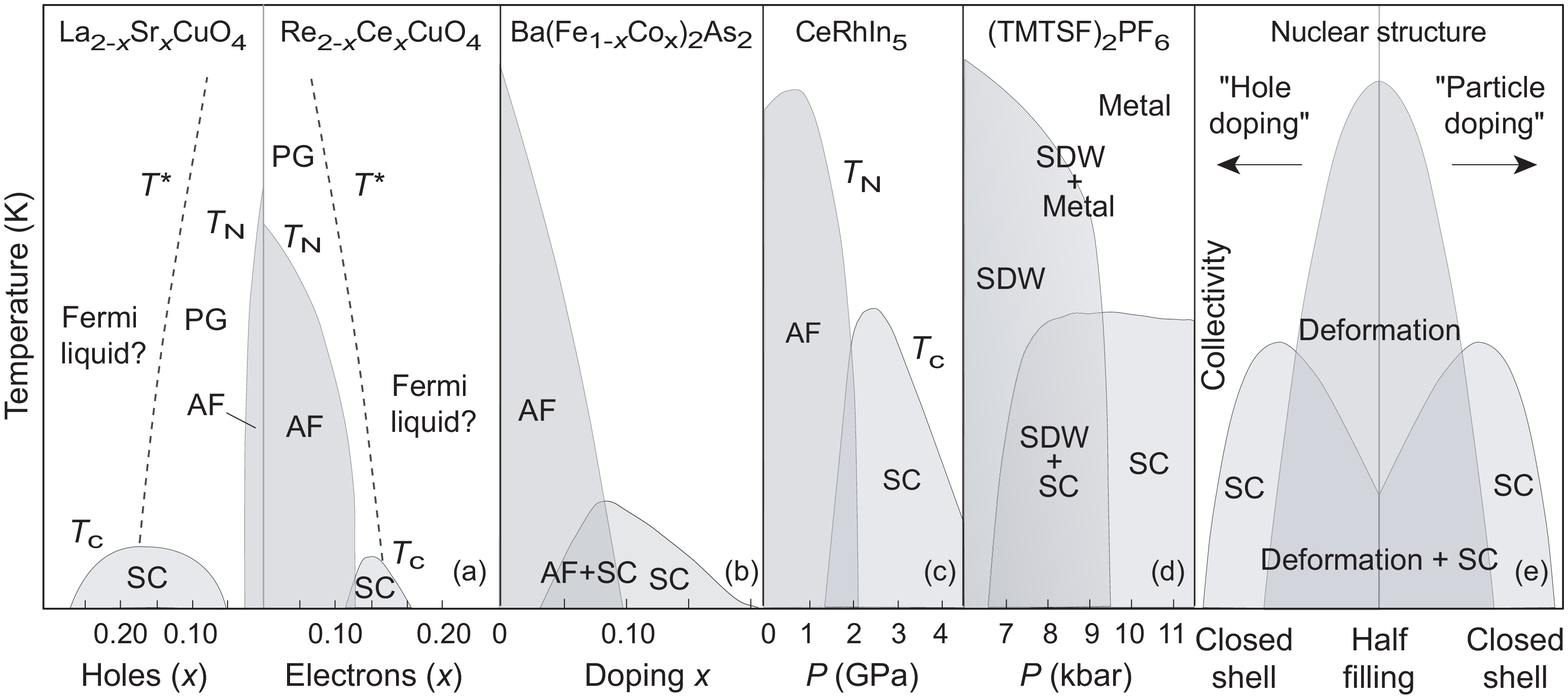}
     {0pt}
     {0pt}
     {0.41}
     {Universality of superconductivity and superfluidity.  (a)~Phase diagram 
for  hole- and electron-doped cuprates \cite{armi2010}. Superconducting (SC), 
antiferromagnetic (AF), and pseudogap (PG) regions are labeled, as are  N\'eel 
($T\tsub N$), SC critical ($T\tsub c$), and PG ($T^*$) temperatures. (b)~Phase 
diagram for Fe-based SC  \cite{fang09}. (c)~Heavy-fermion phase diagram 
\cite{kne09}. (d)~Phase diagram for an organic superconductor (SDW denotes spin 
density waves) \cite{kang10}. (e)~Generic correlation-energy diagram for nuclear 
structure \cite{guid14a}.}
shows that phase diagrams for unconventional SC emergent states in a 
broad range of condensed matter systems and in nuclear structure  are remarkably 
similar, despite completely different microscopic physics (see Refs.\ 
\cite{su4review,guid14a} for further discussion). An even more remarkable 
universality of emergent states  is shown in 
the coherent state energy surfaces displayed in \fig{nuc_SC_graphene_chains}.
\singlefig
     {nuc_SC_graphene_chains}
     {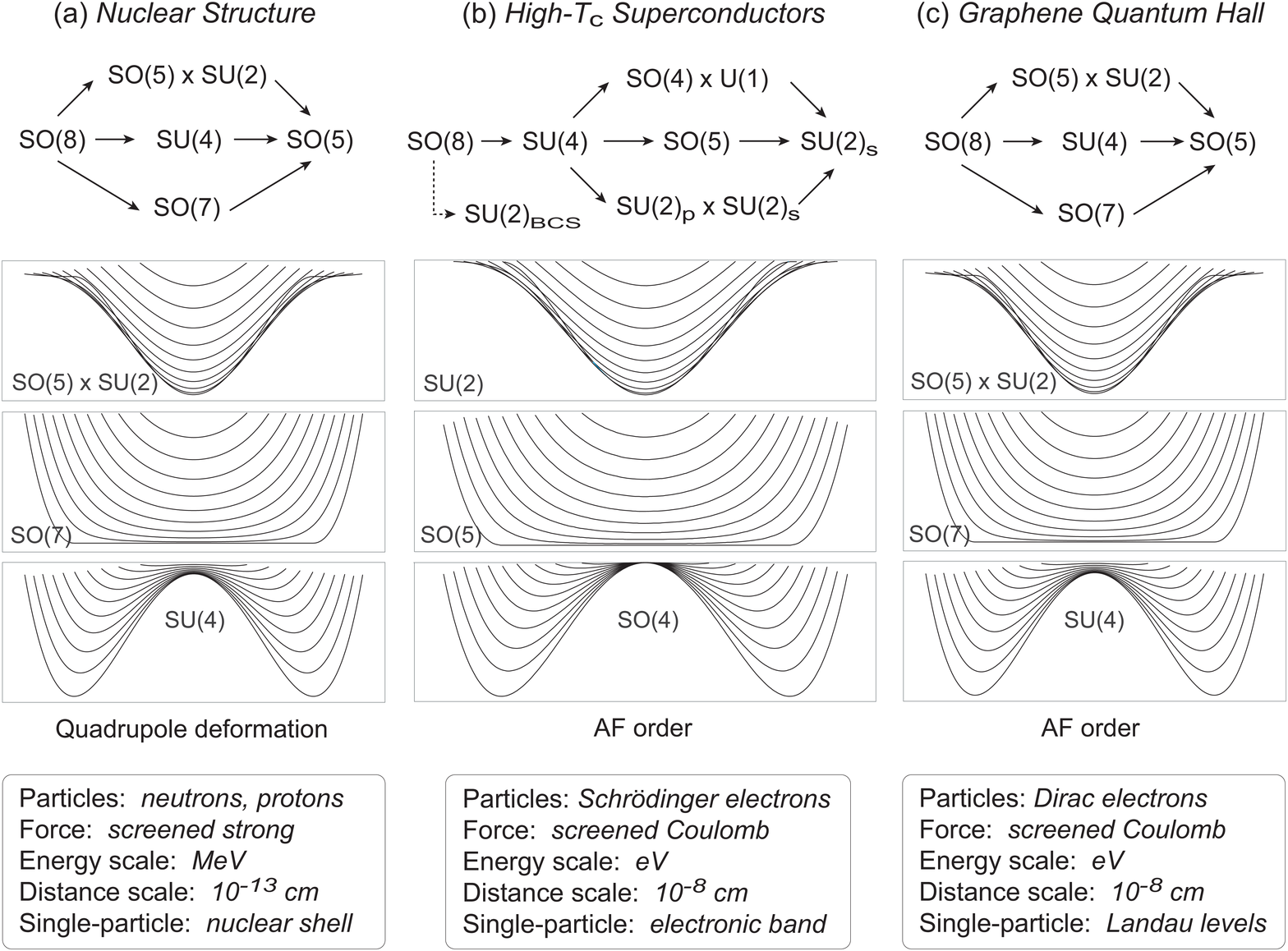}
     {0pt}
     {0pt}
     {0.213}
     {Similarity in the dynamical symmetry chains and the ground coherent state 
energy surfaces for (a)~dynamical symmetry in nuclear structure \cite{wu94}, 
(b)~high-temperature SC \cite{guid2001,lawu03}, and (c)~monolayer graphene in a 
strong magnetic field \cite{wu2016}. The plot contours show total energy as a 
function of an appropriate order parameter, with different curves corresponding 
to a particle number parameter.}
Seemingly different emergent modes:  collective states for atomic nuclei, for 
graphene in a magnetic field, and for cuprate high-temperature SC, give nearly 
identical energy surfaces under a suitable mapping of respective order 
parameters and rescaling of energy. Microscopically these modes differ 
fundamentally but they share a common Hilbert-space truncation to a collective 
subspace dictated by shared Lie algebras that care only about commutators of 
generators, not their microscopic structure \cite{su4review}.

\section{Transition Temperatures for Unconventional Superconductors}
\label{sec:highTcunconventional}

The discussion above shows that \su4 dynamical symmetry describes 
a broad range of cuprate SC properties using a minimum of 
adjustable parameters or adjustable assumptions.  Having established that it 
should be taken seriously, let's  ask what \su4 symmetry has to say about 
$T\tsub c$. The generators of \so4  antiferromagnetism and  \su2 
superconductivity are also generators of \su4.  This implies that  (1)~AF and SC 
compete for the collective Hilbert subspace, and (2)~AF states and SC states are 
related by a rotation within the \su4 group space. As we now explain, this 
implies on fundamental grounds a higher than average $T\tsub c$
for unconventional SC in general, and for cuprate and Fe-based HTSC in 
particular.

A generic reason for higher $T\tsub c$ in unconventional SC is illustrated in 
\fig{collectiveArrowAlignment},%
\singlefig
     {collectiveArrowAlignment}
     {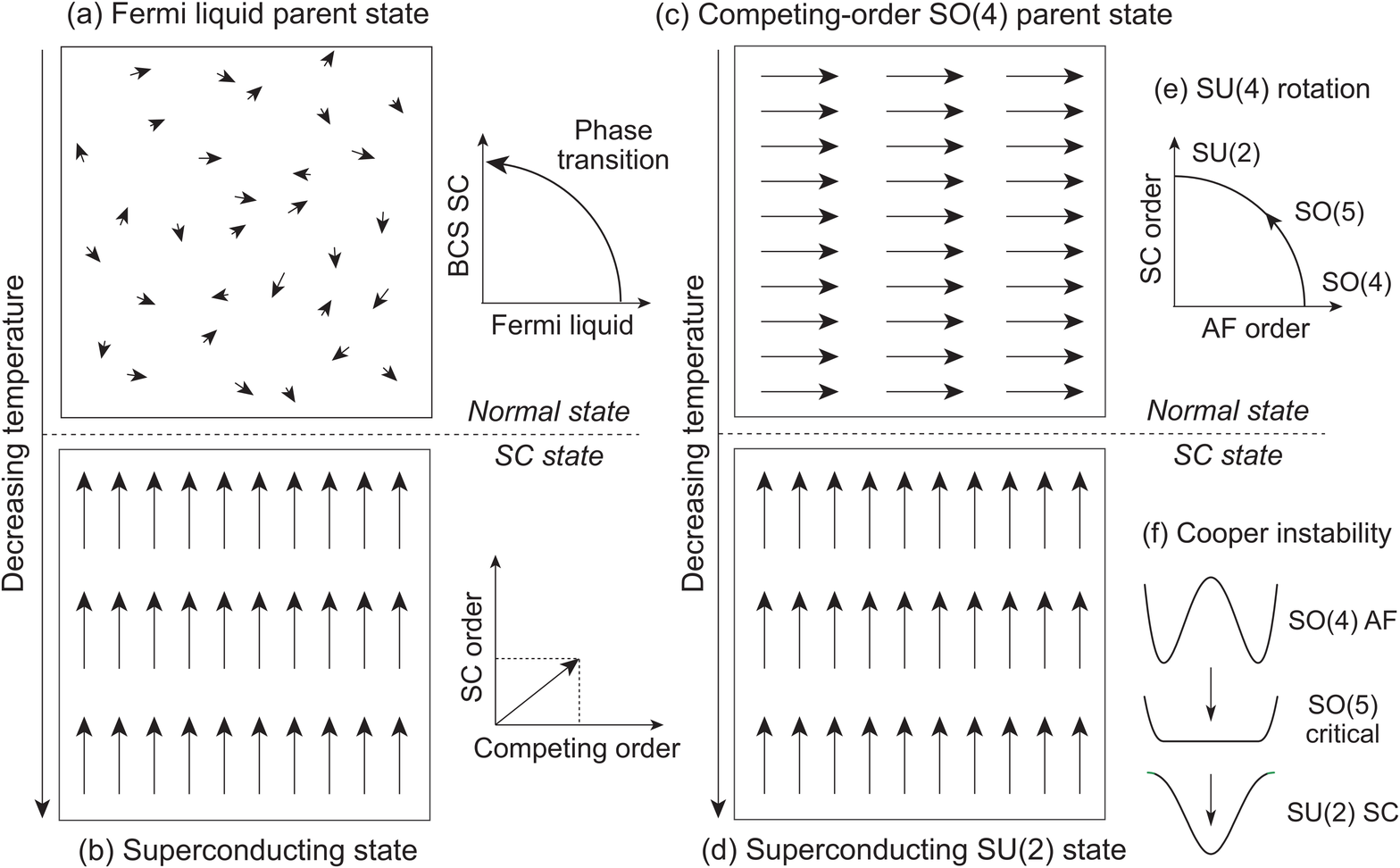}
     {0pt}
     {0pt}
     {0.157}
     {(a)--(b)~Formation of BCS superconductor by lowering the temperature of a 
Fermi liquid through  $T\tsub c$. Direction of vectors indicates relative 
strength of competing order ($x$) and SC ($y$); length indicates total 
\su4 strength.   The SC transition converts a  high-entropy state (a)
into a highly-ordered one (b), implying a low  $T\tsub c$. (c)--(d)~Formation 
of SC from a parent state having order that competes with SC but is related to 
SC by symmetry. This requires imposing SC order (d) on a state (c)
already highly ordered, which can occur at a higher $T\tsub c$ 
because it is a collective rotation in the group space between two low-entropy 
states. (e)~Collective rotation in \su4 group space.  (f)~\su4 Cooper 
instability.}
where the $x$ component of arrows may be viewed as the average competing-order 
matrix element squared and the $y$ component as the average pairing matrix 
element squared. In the conventional BCS case of 
\fig{collectiveArrowAlignment}(a-b) there is no net SC or AF in the initial 
state of \fig{collectiveArrowAlignment}(a) and arrows are small and randomly 
oriented.  To form the superconducting state from the parent state each arrow 
must be lengthened  and ordered vertically in the SC phase transition, as in 
\fig{collectiveArrowAlignment}(b).  This is a transition from a high-entropy 
initial state to a highly-ordered final state, implying that it requires a 
corresponding low temperature to implement. In the competing order case the 
initial state of \fig{collectiveArrowAlignment}(c) is already ordered such that 
a simple collective rotation can produce the SC state in 
\fig{collectiveArrowAlignment}(d). This is the general case when SC and the 
competing order are unified by symmetry.  In essence, the collective vectors 
already exist as a highly-ordered configuration in the parent state with a 
length proportional to the \su4 Casimir expectation value, but they point in the 
AF direction.  To produce a superconductor from a parent AF state, they need 
only be rotated uniformly to point in the SC direction; 
\fig{collectiveArrowAlignment}(e) illustrates. Thus, if competing order and  
superconductivity are related by  symmetry the  parent state can 
``pre-condition'' the phase transition, allowing it to occur at higher  $T\tsub 
c$ because the low-entropy competing-order state can be rotated collectively 
into the low-entropy superconducting state.  

We have shown that the \su4 model exhibits a generalized Cooper instability 
whereby the AF Mott insulator state reorganizes spontaneously into a 
superconductor when it is perturbed by adding  holes \cite{guid10}. The 
collective rotation in the \su4 space indicated  in 
\fig{collectiveArrowAlignment}(c-e) is a schematic representation of this 
generalized Cooper instability, which can occur spontaneously if there is no 
barrier to the rotation. The $\su4\supset \so5$ critical dynamical symmetry 
exhibits such a property, as illustrated in \fig{collectiveArrowAlignment}(f). 
At low doping the energy surface implies degenerate AF and SC ground states with 
no energy barrier between them [see \fig{eSurfacesAndCompare}(b) for lattice 
occupation fraction of one].  This suggests that the AF and SC phases can be 
connected by a sequence of infinitesimal \su4 rotations through intermediate 
states having different mixtures of AF and SC order that are degenerate in 
energy.

These entropy arguments are equivalently information arguments.   Figure 
\ref{fig:collectiveArrowAlignment}(d) is obtained from 
\fig{collectiveArrowAlignment}(c) by collective rotation, which requires 
specification of a single angle. Conversely, there is no order in the parent 
state of \fig{collectiveArrowAlignment}(a) and each arrow must be lengthened and 
oriented individually to give \fig{collectiveArrowAlignment}(b), which requires 
supplying much more information.  Thus the ordering necessary to condense the SC 
state  is much larger in Figs.\  \ref{fig:collectiveArrowAlignment}(a-b) than in 
Figs.\  \ref{fig:collectiveArrowAlignment}(c-d). The information argument also 
makes clear the essential difference between competing collective modes that are 
independent and those related by a symmetry.  In the former case a large amount 
of information is required to change the competing-order state into the SC state 
because they are unrelated: the symmetry of the parent state must be destroyed 
and the SC symmetry then constructed from the pieces.  In the latter case the 
symmetry already encodes the relationship between the two modes; hence only a 
small amount of information is required to produce the SC state from the 
competing-order state.

Our discussion has emphasized cuprate SC examples but  applies generally to 
unconventional SC.  The crucial physics of unconventional superconductivity lies 
in the competition of other collective modes with pairing, and possible Coulomb 
repulsion effects.  These polarize the pairing interaction and alter the 
geometry of the pairing formfactor, but that  is {\em symptomatic;}  the 
essential physics lies in the competing order, not the formfactor. Thus, the 
reason for high-$T\tsub c$ proposed here should be operative in all 
unconventional SC, leading to the often abnormally high $T\tsub c$ (on an 
appropriate scale) seen for 
cuprates, Fe-based SC, heavy-fermion SC, and other unconventional 
superconductors, by virtue of universal symmetry arguments depending only 
parametrically on microscopic details like pairing geometry.

This work was partially supported by the National Key Program for S\&T Research 
and Development (Grant No. 2016YFA0400501), and by LightCone Interactive LLC.  
L. W. acknowledges grant support from the Basque Country Government (Grant No. 
IT986-16), and by PGC2018-101355-B-I00 (MCIU/AEI/FEDER,UE).

\end{document}